\documentclass{iau}

\usepackage{amsmath}
\usepackage{graphicx}
\usepackage{multirow}

\begin{document}

\lefttitle{Kostadinka Koleva et al.}
\righttitle{Filament eruption deflection and associated CMEs}

\jnlPage{1}{7}
\jnlDoiYr{2024}
\doival{10.1017/xxxxx}
\volno{388}
\pubYr{2024}
\journaltitle{Solar and Stellar Coronal Mass Ejections}

\aopheadtitle{Proceedings of the IAU Symposium}
\editors{N. Gopalswamy,  O. Malandraki, A. Vidotto \&  W. Manchester, eds.}

\title{Filament eruption deflection and associated CMEs}

\author{K. Koleva$^{1}$, R. Chandra$^{2}$, P. Duchlev$^{1}$, P. Devi$^{2}$}
\affiliation{$^{1}$Space Research and Technology Institute, Bulgarian Academy of Sciences, Sofia, Bulgaria\\
	$^{2}$Department of Physics, DSB Campus, Kumaun University, Nainital
}

%\author{Cambridge Author1}
%\affiliation{Cambridge University Press}

\begin{abstract}
We present the observations of a quiescent filament eruption and its deflection from the radial direction. The event occurred in the southern solar hemisphere on 2021 May 9 and was observed by the Atmospheric Imaging Assembly (AIA) on board the Solar Dynamics Observatory (SDO), by the STEREO -A Observatory and GONG. Part of the filament erupted in the west direction, while major part of the filament deviated towards east direction. LASCO observed a very weak CME towards the west direction where it faded quickly. Moreover, the eruption was associated with CME observed by STEREO A COR1 and COR2. Our observations provide the evidence that the filament eruption was highly non-radial in nature.
\end{abstract}


\maketitle

\section{Introduction}
Solar filament eruptions are one of the most studied phenomena in the field of solar physics. Filaments can erupt fully or partially and these erupted material can be observed in form of coronal mass ejections (CMEs) (\cite{1976SoPh...48..159W, 1979SoPh...61..201M,2000AdSpR..25.1851G, 2003AdSpR..31..869G, 2013AdSpR..51.1967S, 2021EP&S...73...58S}  , and references cited therein).  There are cases of the deflection of CME during its propagation near to Sun or in the interplanetary space (for example:
 \cite{2012ApJ...744...66Z, 2024ApJ...966...22K}. The study of the deflection of filament eruption and CMEs can be a crucial study to understand the geoeffectiveness of solar eruptions.

In this work we present the study of a sigmoid filament eruption that occurred in the  southern solar hemisphere on 2021 May 9. The kinematics and morphology of the event were investigated by \cite{2021simi.conf...34K}.  Here we focused on the associated CME and its offset from the radial propagation.

The used data and observations from two points of view are introduced in Section 2. Section 3 provides our interpretation of the observations.

\section{Data and Observations}

The prominence eruption occurred at the southern solar hemisphere between 9:30 and 11:00 UT on 2021 May 9. 

For the present study, we used the data from the Atmospheric Imaging Assembly \cite[AIA:][]{2012SoPh..275...17L} on board the Solar Dynamics Observatory \cite[SDO:][]{2012SoPh..275....3P} in 304~\AA{}. For a different viewpoint of the event, we used data from the Sun Earth Connection and Heliospheric Investigations \cite[SECCHI:][]{2008SSRv..136...67H} instruments on board the Solar Terrestrial Relations Observatory Ahead \cite[STEREO-A:][]{2008SSRv..136....5K} spacecraft. More specifically, data from the Extreme Ultraviolet Imager (EUVI) in the 304 \AA{} channel, the STEREO-A COR1 inner coronagraph \citep{2003SPIE.4853....1T}, and the COR2 outer coronagraph are used in the study.

We used also data obtained by the XRT \citep{2007SoPh..243...63G} onboard HINODE \citep{2007SoPh..243....3K} to follow the filament evolution one day before the eruption.

\begin{figure}
  	\begin{center}
  	\includegraphics[scale=.6]{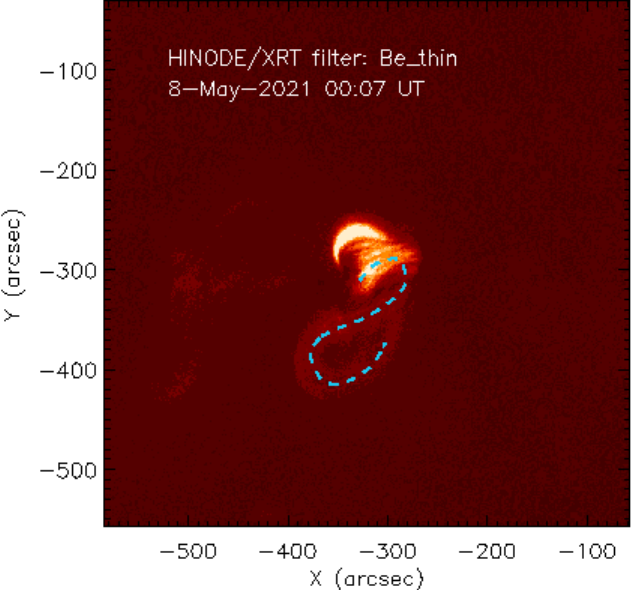}
  	\caption{HINODE/XRT observation on 2021 May, 8 at 00:07 UT. The blue dotted line traces the sigmoid structure.}
  	\label{f1}
  	\end{center}
  \end{figure}

Before the eruption, the sigmoid filament was located in a plage region close to the disk center and laid along the S-shaped magnetic polarity inversion line. A weak X-ray sigmoid was observed in Hinode/XRT Al and Be filters along the filament channel one day before the eruption. In Figure~\ref{f1} the sigmoid on 8 May 2024, observed  in the Be filter image of HINODE/XRT instrument is presented.

\subsection{AIA/SDO FOV}

The eruption began at around 09:30 UT in the AIA field-of-view (FOV). Part
of the filament erupted in the west direction, while major part of the filament deviated towards east direction. Before deviation from the radial direction, the erupted filament undergoes a twisting motion.

  \begin{figure}
	\includegraphics[scale=.7]{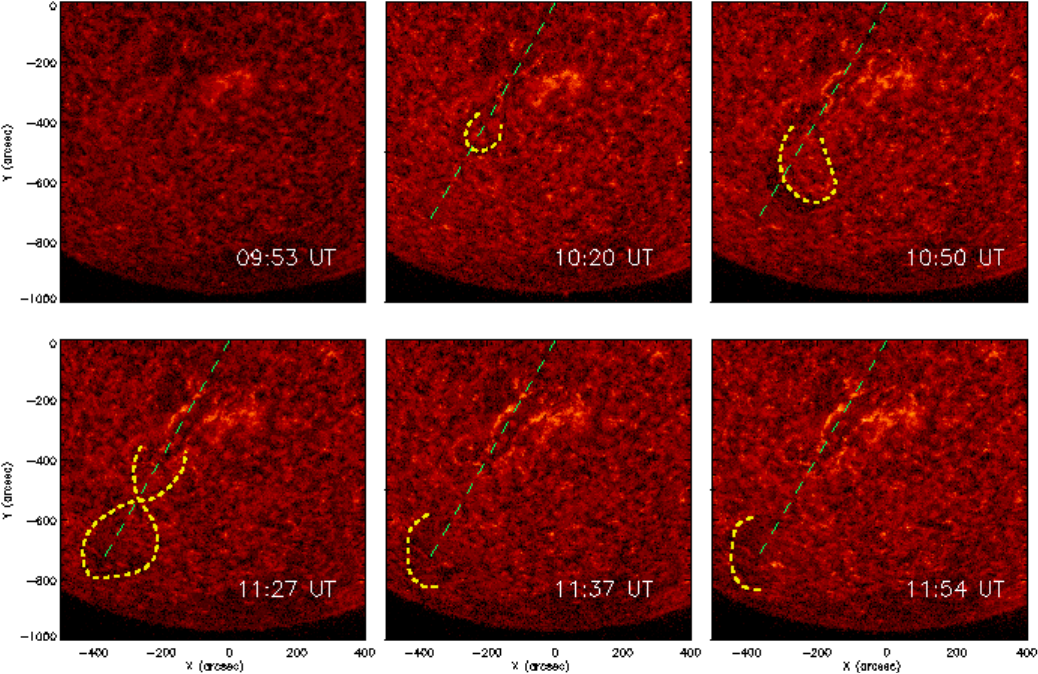}
	\caption{The eruption evolution as seen by AIA/SDO in 304 \AA{} channel. The green dotted line shows the radial path through the filament loop center at 10:20 UT.}
	\label{f2}
\end{figure}

The eruption evolution is shown in Figure~\ref{f2} in AIA 304 \AA{} channel. The radial path through the filament loop center at 10:20 UT is indicated by the green dotted line. The filament's loop is traced by the yellow dotted line. The eastward deflection is well observed.

\subsection{EUVI/STEREO A FOV}
\begin{figure}
  	\begin{center}
  	\includegraphics[scale=.6]{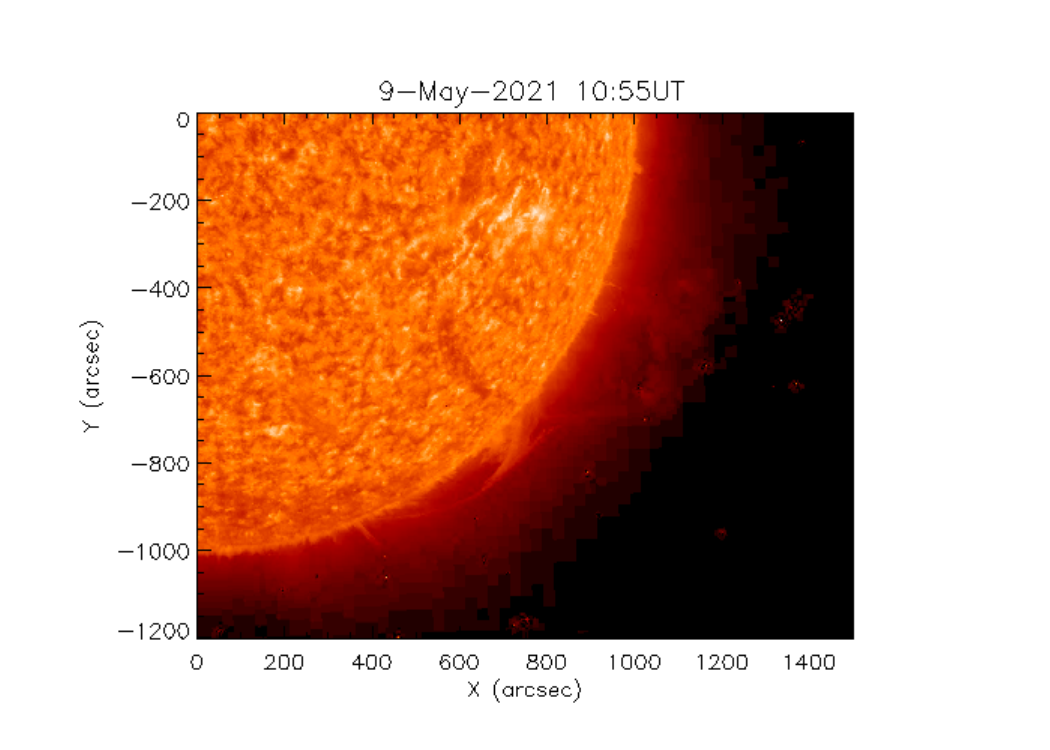}
  	\caption{Filament eruption observed by EUVI/STEREO A in 304 \AA{} channel at 10:55 UT.}
  	\label{f3}
  	\end{center}
  \end{figure}

The eruption was observed also by STEREO Ahead spacecraft close to the bottom right limb. At the time of observation the separation angle of the STEREO-A spacecraft  with Earth was $52^\circ$. The filament eruption as observed by EUVI on board the STEREO A spacecraft is shown in Figure~\ref{f3}.

\begin{figure}
	\includegraphics[scale=.4]{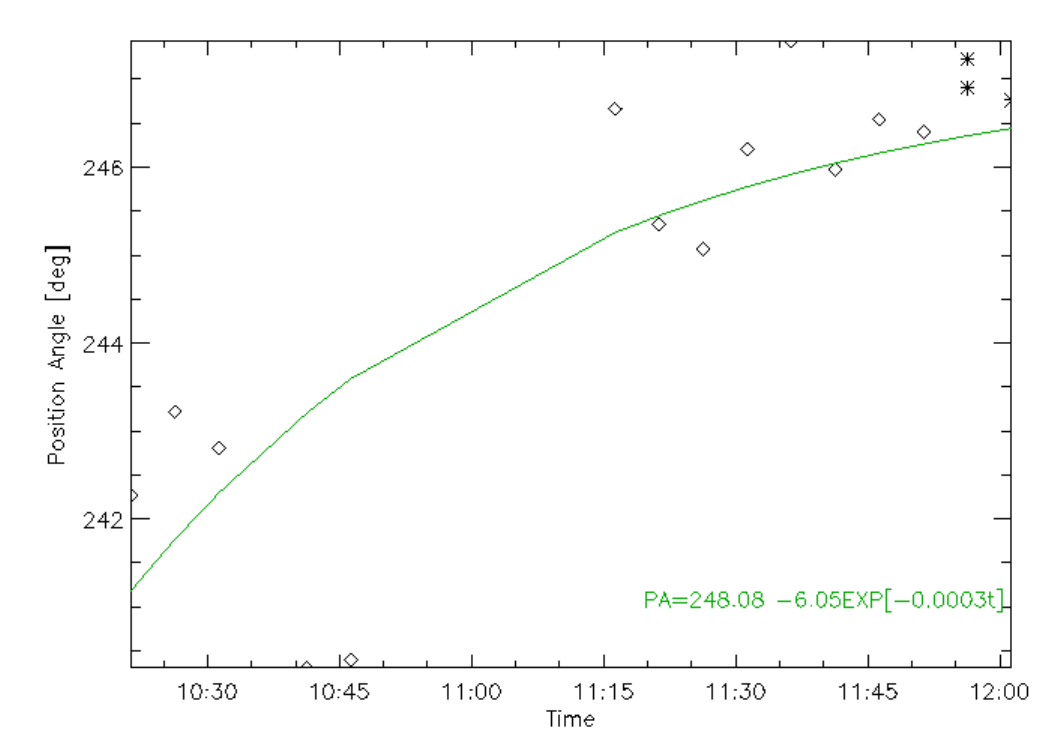}
		\includegraphics[scale=.4]{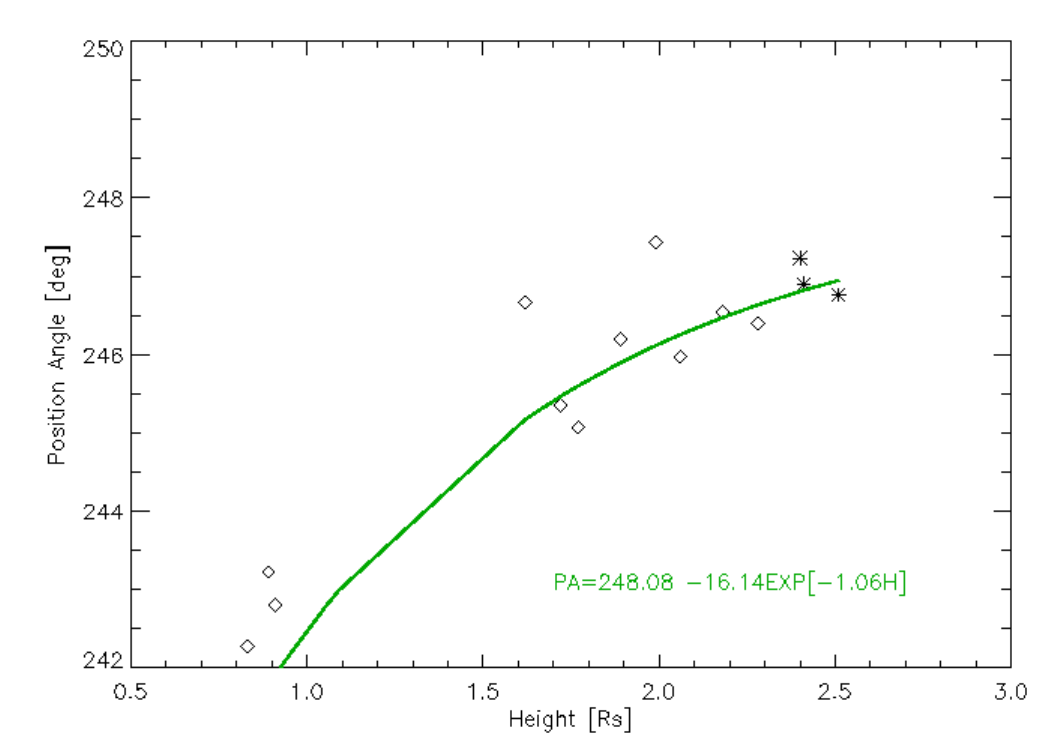}
	\caption{\textit{left:} Variation of the position angle of CME core as a function of time.
		Diamonds and stars represent measurements from COR1 and COR2
		coronagraphs of STEREO A , respectively.\\
		\textit{right:} Prominence Position angle (PA) at various heliocentric distances.
		The solid line is the fit to the data points.}
	\label{f4}
\end{figure}

In the COR1 FOV the prominence eruption starts at 10:21 UT at Position angle (PA) $242^\circ$ and left COR1 FOV at 12:01~UT at PA $247^\circ$ (see Figure~\ref{f4}, left panel). The deflection stopped at about 1.94 solar radii (Figure~\ref{f4}, right panel). After that the core continues its propagation in C2 FOV without any visible deflection.

\section{Results}

\begin{figure}
  	\begin{center}
  	\includegraphics[scale=.35]{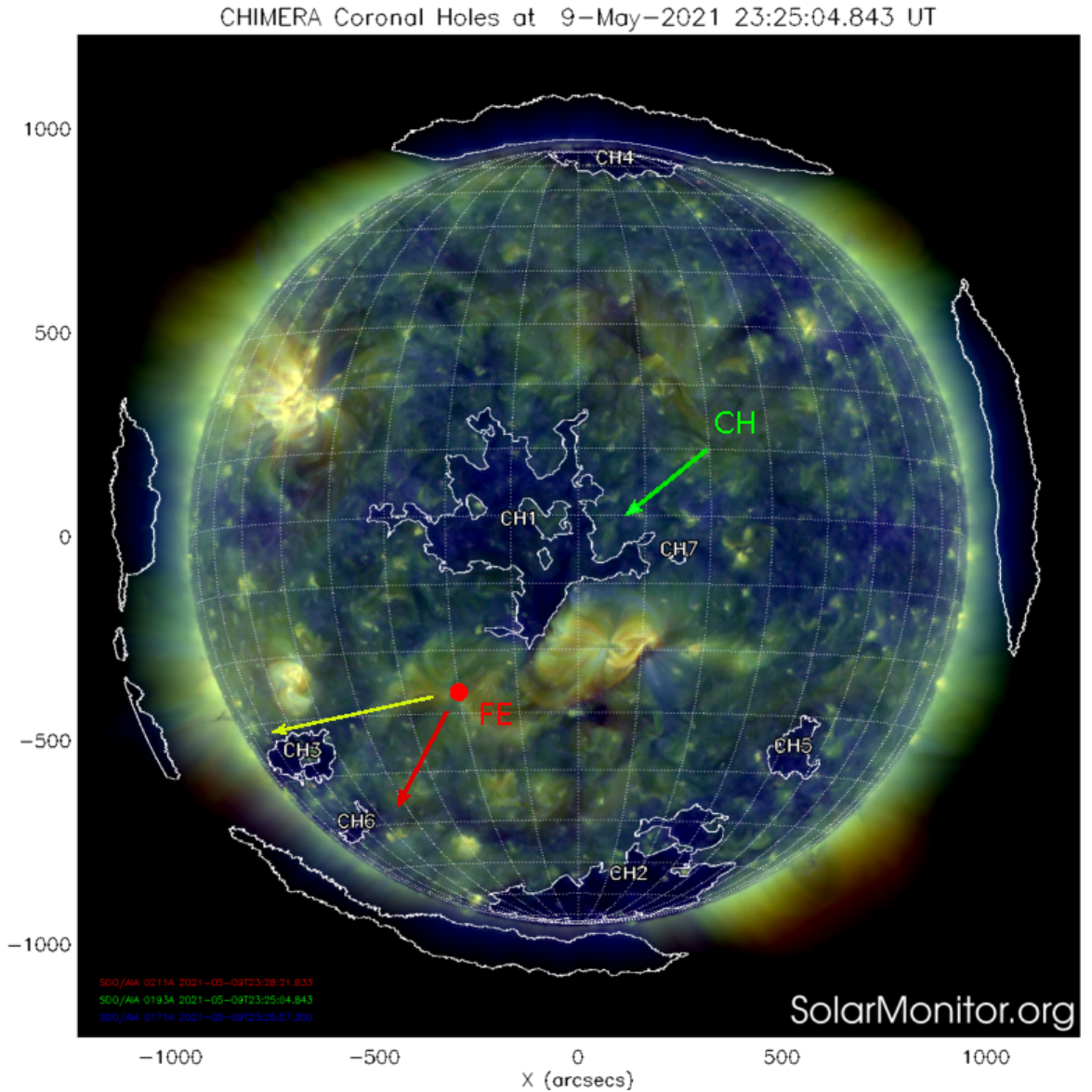}
  	\caption{The composite image of three AIA channels: 211~\AA{}, 193~\AA{} and 171~\AA{} at 9 May 2021. The eruption direction at 10:20 UT and 11:54 UT, is indicated by the red and yellow arrows, respectively.}
  	\label{f5}
  	\end{center}
  \end{figure}

We interpret our result as follows: part of the filament went into west direction in the AIA FOV 
and appeared as a weak CME in LASCO C2 coronagraph at around 13:25~UT. The major part of the
erupted filament interact with the southern situated filament 
and deviated in the eastern direction. The eruption starts at a position angle at about $153^\circ$. After deviation the position angle of the  prominence loop center was about $145^\circ $.
In Figure~\ref{f5} the eruption direction at 10:20 UT is indicated by the red arrow, while the yellow marks the eruption direction at 11:54~UT. The angle between the two arrows, which represents the deflection angle was approximately $8^\circ$. The deflected filament part may be directed away from the earth direction, and it was observed as a CME by COR~1 and COR~2 instrument.

\section*{Acknowledgements}
K.K. acknowledges the International Astronomical Union grant.
K.K. and P.D. acknowledge the support of Bulgarian Science Fund under grand No KP-06-H44/2 27.11.2020.
P.D. thanks the CSIR, New Delhi for providing the research fellowship.

\end{document}